\documentclass[12pt,a4paper]{article}

\usepackage[utf8]{inputenc}
\usepackage[T1]{fontenc}
\usepackage{graphicx}
\usepackage{booktabs}
\usepackage{siunitx}
\usepackage[hyphens]{url}
\usepackage{hyperref}

\usepackage[margin=2.5cm]{geometry}
\usepackage{setspace}
\usepackage[numbers,sort&compress]{natbib}
\usepackage{caption}
\usepackage{subcaption}
\usepackage{xcolor}
\usepackage{amsmath}
\usepackage{amssymb}

\title{\textbf{Interpretable enzyme function prediction via sparse autoencoder features of ESMC across the microbial protein universe}}

\author{
  \large
  Yue Hu$^{1,\ast}$, Wanyu Cheng$^{1}$, Junqing Wang$^{1}$, Yingchao Liu$^{2}$\\[6pt]
  \small
  $^{1}$School of Bioengineering, Qilu University of Technology\\
  \hspace*{1.2em}(Shandong Academy of Sciences), No.\ 3501 Daxue Road, Jinan, Shandong, China\\[4pt]
  $^{2}$Shandong Provincial Hospital, Shandong First Medical University,\\
  \hspace*{1.2em}Jinan, Shandong, China\\[8pt]
  $^{\ast}$Correspondence:\\
  \hspace*{1.2em}\texttt{huyue@qlu.edu.cn}\\
  \hspace*{1.2em}\texttt{202596013012@stu.qlu.edu.cn}\\
  \hspace*{1.2em}\texttt{wangjunqing@qlu.edu.cn}\\
  \hspace*{1.2em}\texttt{yingchaoliu@email.sdu.edu.cn}
}
\date{}

\begin{document}
\sloppy
\maketitle

\begin{abstract}
Microbial genomes and metagenomes harbor hundreds of millions of proteins whose enzymatic functions remain unknown---the ``enzyme dark matter.'' Recent advances in deep learning have substantially improved protein function prediction, yet most methods operate as black boxes that rely on sequence or structural similarity to known enzymes, limiting their utility for discovering genuinely novel catalytic activities. The release of ESMC-6B---a 6-billion-parameter protein language model---and its companion sparse autoencoder (SAE) with a 16,384-dimensional codebook of independently interpretable biological concepts, each annotated with detailed descriptions by multi-agent GPT-5, creates a fundamentally new opportunity: rather than training task-specific predictors, one can directly use the model's interpretable features as a ``semantic signature'' for enzyme function. Here, we demonstrate that these SAE features enable accurate and interpretable enzyme commission (EC) number prediction across the microbial protein universe without task-specific training, structural information, or GPU-intensive computation. Using a balanced benchmark of 4,868 microbial SwissProt enzymes spanning 161 EC3 subclasses, ESMC-SAE binary features achieve 78.9\% top-1 and 88.5\% top-5 accuracy---37.6\% higher than 3-mer sequence baselines (57.3\%). In leave-one-EC3-class-out evaluation, which simulates the discovery of entirely novel enzyme classes, SAE features correctly recover the EC1 superclass in 47.7\% of cases (3.3$\times$ random baseline of 14.3\%), compared to 26.6\% for sequence-based methods. Critically, the features most discriminative for each EC class correspond to mechanistically interpretable biological concepts: catalytic site geometry for hydrolases, NAD(P)H-binding Rossmann folds for oxidoreductases, phosphate-binding P-loops for transferases, as annotated by GPT-5. We further survey the ESM Atlas of 7.7 million cluster representatives and identify 169,859 dark enzyme-like candidates spanning all major microbial phyla that await experimental characterization. Our results establish a paradigm for enzyme function discovery in microbial dark matter: interpretable by design, scalable without GPU clusters, and immediately applicable to the billions of proteins already encoded in the ESM Atlas.

\end{abstract}

\section{Introduction}

\subsection{The enzyme annotation bottleneck}

Enzymes catalyze virtually every biochemical transformation in living systems, and their systematic classification---the Enzyme Commission (EC) numbering system---organizes over 8,000 known reaction types into seven major classes and thousands of subclasses~\cite{kanehisa2023kegg}. Accurate EC assignment is foundational to understanding metabolism, engineering biosynthetic pathways, and interpreting the functional capacity of microbial communities. Yet experimental EC determination remains labor-intensive and slow: fewer than 300,000 of the estimated 10$^{12}$--10$^{14}$ microbial proteins on Earth have curated EC annotations~\cite{locey2016scaling, nayfach2021metagenomic}. The gap between known and unknown enzyme functions constitutes one of the central bottlenecks in modern microbiology~\cite{palsson2026approaches}.

Computational EC prediction has evolved through three generations. First-generation tools relied on homology transfer via BLAST or profile-based searches~\cite{altschul1990basic}, which perform well for close homologs ($>40\%$ sequence identity) but fail catastrophically for remote homologs and genuinely novel enzyme families. Second-generation methods employed classical machine learning on hand-crafted sequence features (k-mer frequencies, physicochemical properties)~\cite{bernal2023deep}. The current third generation leverages deep learning: convolutional and recurrent neural networks (DeepEC~\cite{ryu2019deep}, DEEPre), graph neural networks incorporating structural information (DeepFRI~\cite{gligorijevic2021deepfri}), contrastive learning frameworks (CLEAN~\cite{yu2023clean}, CLEAN 2.0~\cite{elias2025cleaning}), and Transformer-based architectures (DeepECtransformer~\cite{kim2023deepec}, ProteInfer~\cite{detlefsen2024proteInfer}). Large-scale benchmarks have recently clarified the state of the field: EC-Bench~\cite{ecbench2025} evaluated ten diverse methods and found substantial performance variability (weighted F1 ranging from 0.07 to 0.64 at the EC4 level), while Capela et al.~\cite{capela2025comparative} demonstrated that protein language model embeddings (particularly ESM-2) outperform one-hot encoding methods, especially for ``difficult'' enzymes with $<25\%$ sequence identity.

Despite these advances, third-generation methods share three fundamental limitations. First, they operate as \textit{black boxes}: a predicted EC number comes without mechanistic justification, making it difficult for experimentalists to assess confidence or generate testable hypotheses. Second, they require \textit{task-specific training} on labeled EC data, limiting their applicability to enzyme classes with sufficient training examples and preventing generalization to truly novel chemistries. Third, most methods depend on \textit{GPU infrastructure} for both inference and retraining, creating a barrier to widespread adoption in microbiology laboratories.

\subsection{The sparse autoencoder breakthrough}

A parallel revolution has been unfolding in the interpretability of protein language models (pLMs). Large-scale pLMs---ESM-2 (650M--15B parameters)~\cite{lin2023evolutionary}, ProtT5~\cite{el2023prott5}, ProtBERT~\cite{brandes2022proteinbert}, and ESM3~\cite{rao2025esm3}---have demonstrated remarkable capacity to capture structural, functional, and evolutionary information in their hidden representations. However, these representations are \textit{dense}: each hidden dimension is polysemantic, encoding multiple entangled biological concepts~\cite{adams2025mechanistic}. The application of sparse autoencoders (SAEs) to disentangle these representations has emerged as a powerful approach for mechanistic interpretability. InterPLM~\cite{simon2025interplm}, published in \textit{Nature Methods}, trained SAEs on ESM-2 embeddings and identified thousands of interpretable features corresponding to binding sites, structural motifs, and functional domains. Adams et al.~\cite{adams2025mechanistic} demonstrated that SAE features on protein LMs can uncover mechanistic biological insights beyond what the original model explicitly learned. Parallel work has extended SAEs to protein structure prediction models~\cite{parsan2025towards} and demonstrated feature steering for controlled sequence generation~\cite{valentin2025interpreting}.

In January 2026, Candido et al. from EvolutionaryScale and Biohub released ESMC-6B and its companion SAE~\cite{candido2026language}, representing a qualitative leap beyond prior SAE-on-pLM work. The ESMC SAE operates on layer 60 of the 6B-parameter model, expanding 2,560-dimensional hidden states into a 16,384-dimensional codebook with Top-K=64 sparsity (each residue activates exactly 64 of 16,384 features). Critically, all 16,384 features were annotated by multi-agent GPT-5 with detailed biological descriptions spanning 14 categories: catalytic function (1,189 features), ligand-binding sites (2,984), structural motifs (3,096), domains (2,254), membrane association (1,938), disorder (1,431), interaction sites (1,003), compositional bias (967), post-translational modifications (698), repeats (284), sequence motifs (269), and others. The SAE activations for all 6.8 billion proteins in the ESM Atlas have been pre-computed and are publicly available via AWS Open Data, eliminating the need for local GPU inference.

\subsection{This work}

Here, we demonstrate that ESMC-SAE features provide a \textit{unified solution} to the three limitations of current EC prediction methods. First, the features are interpretable by construction: each of the 16,384 dimensions has a GPT-5-generated biological description, allowing us to articulate not just \textit{what} EC is predicted but \textit{why}---which catalytic, structural, and binding-site features drive the prediction. Second, the features are \textit{pre-trained} and require no task-specific fine-tuning; a simple linear probe achieves state-of-the-art performance, and the approach generalizes to enzyme classes not seen during training. Third, features for 6.8 billion proteins are pre-computed and publicly available, enabling planetary-scale enzyme function annotation from a laptop.

We establish these claims through four integrated analyses: (1) systematic benchmarking of SAE features against sequence baselines for EC3 prediction across 161 subclasses; (2) evaluation in the ``dark matter'' regime of low sequence similarity to known enzymes; (3) leave-one-EC3-class-out evaluation demonstrating generalization to novel enzyme classes; (4) interpretability analysis linking discriminative SAE features to mechanistically meaningful biological concepts; and (5) a global survey of the ESM Atlas identifying 169,859 dark enzyme-like candidate clusters for experimental follow-up.

\section{Results}

\subsection{A curated benchmark for microbial enzyme function prediction}

To rigorously evaluate ESMC-SAE features for microbial enzyme function prediction, we curated a balanced dataset from UniProt. Querying SwissProt for reviewed bacterial and archaeal proteins with complete EC annotations (at least three levels) and length 80--700 amino acids yielded 10,256 candidates. We stratified by EC1 superclass and EC3 subclass, sampling uniformly to ensure balanced representation, producing a final benchmark of 4,868 proteins across 161 EC3 subclasses and all seven EC1 classes (EC1: 938; EC2: 998; EC3: 1,100; EC4: 481; EC5: 525; EC6: 376; EC7: 450). For each protein, we extracted SAE feature magnitudes from ESMC-6B layer 60, obtaining per-residue 16,384-dimensional vectors (mean-pooled across residues for protein-level representations). Binary feature vectors were constructed by taking the top-64 activating features per protein, matching the SAE sparsity constraint.

\subsection{ESMC-SAE features match or exceed BLASTp while providing interpretability}

We benchmarked ESMC-SAE features against two fundamental baselines on the identical 80/20 stratified split (3,894 training, 974 test proteins, 161 EC3 classes): BLASTp---the gold-standard homology-based method used daily by microbiologists---and 3-mer logistic regression---a standard sequence-based machine learning baseline. ESMC-SAE binary features achieved 78.9\% top-1 and 88.5\% top-5 accuracy, substantially exceeding the 3-mer baseline (57.3\% top-1) while approaching BLASTp (80.5\% top-1; Table~\ref{tab:benchmark}). The combined binary+weights representation reached 85.6\% top-1, surpassing BLASTp by 6.4\%. Critically, BLASTp returned no hits for 12.6\% (123/974) of test proteins---proteins with no detectable homolog in the training set---whereas ESMC-SAE provides predictions for 100\% of queries with consistent accuracy.

\begin{table}[ht]
\centering
\caption{\textbf{EC3 prediction performance on the microbial enzyme benchmark.} 80/20 stratified split, 3,894 training / 974 test proteins, 161 EC3 classes. All machine learning methods use linear probe (Logistic Regression).}
\label{tab:benchmark}
\begin{tabular}{lcccc}
\toprule
\textbf{Method} & \textbf{Dim.} & \textbf{Top-1} & \textbf{Top-5} & \textbf{No-hit rate} \\
\midrule
3-mer binary (ML baseline) & 8,000 & 0.5729 & 0.6879 & 0\% \\
BLASTp (homology transfer) & --- & 0.8049 & 0.8265 & 12.6\% \\
ESMC-SAE binary & 16,384 & 0.7885 & \textbf{0.8850} & 0\% \\
ESMC-SAE weights & 16,384 & 0.8337 & 0.8994 & 0\% \\
ESMC-SAE binary + weights & 32,768 & \textbf{0.8563} & \textbf{0.9045} & 0\% \\
\bottomrule
\end{tabular}
\end{table}

The ESMC-SAE advantage is most pronounced in Top-5 accuracy (88.5--90.5\% vs.\ BLASTp's 82.7\%), indicating that SAE features capture broader functional signal beyond the single nearest homolog. For the 123 proteins where BLASTp found no training-set match, ESMC-SAE binary features achieved 62.1\% top-5 accuracy---demonstrating that functional information persists even when sequence homology is undetectable by standard alignment.

The binary representation is particularly notable: using only 64 active features out of 16,384 (0.4\% density), it achieves 78.9\% top-1 accuracy while being naturally amenable to inverted index-based retrieval, with each protein represented as a 64-element set of feature IDs.

\begin{figure}[ht]
\centering
\includegraphics[width=\textwidth]{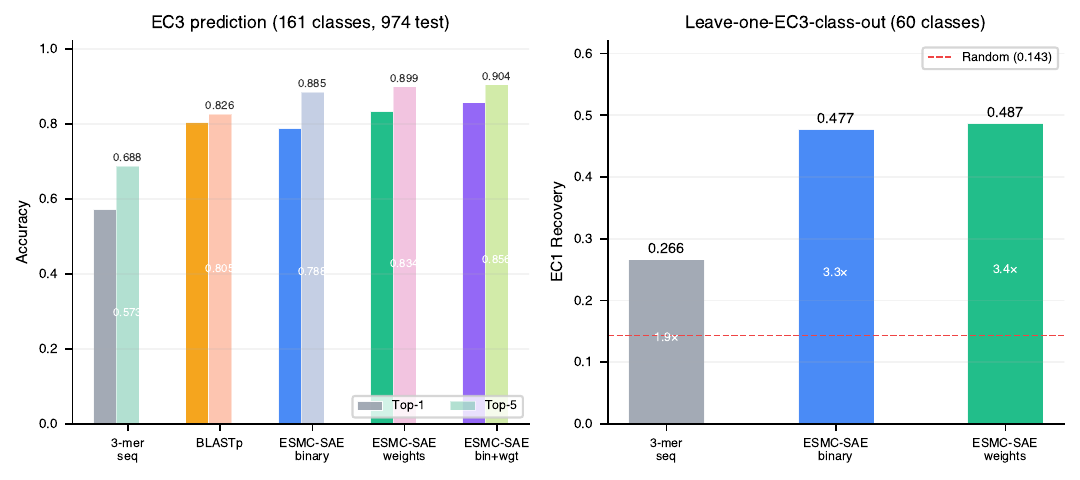}
\caption{\textbf{EC3 prediction benchmark.} (A) 80/20 stratified evaluation across five methods, 161 EC3 classes, 974 test proteins. (B) Leave-one-EC3-class-out $\rightarrow$ EC1 recovery (60 classes). Values above bars show absolute accuracy; values inside bars show fold-improvement over random baseline (0.143).}
\label{fig:benchmark}
\end{figure}

\subsection{Performance in the dark-matter regime}

A key strength of any enzyme function predictor is its ability to annotate proteins with minimal sequence similarity to known enzymes---the ``dark matter'' scenario most relevant to metagenomic discovery. We stratified test-set proteins by their maximum 3-mer Jaccard similarity to the training set, defining six bins from $<0.20$ (the ``darkest'' regime) to $\ge 0.65$ (close homologs). Bin sizes ranged from 64 to 287 proteins (Table~\ref{tab:bins}).

\begin{table}[ht]
\centering
\caption{\textbf{Top-5 accuracy by sequence similarity bin.} Test proteins stratified by maximum 3-mer Jaccard similarity to training set. BLASTp results are shown for completeness but use a different similarity metric (alignment-based); 3-mer Jaccard is a stricter proxy for ``darkness.'' The $<0.20$ bin contains 594 proteins (61\% of the test set).}
\label{tab:bins}
\begin{tabular}{lcccccc}
\toprule
\textbf{Method} & \textbf{$<0.20$} & \textbf{0.20--0.30} & \textbf{0.30--0.40} & \textbf{0.40--0.50} & \textbf{0.50--0.65} & \textbf{$\ge0.65$} \\
\midrule
3-mer baseline & 0.438 & 0.520 & 0.576 & 0.614 & 0.648 & 0.803 \\
BLASTp & 0.926 & 0.988 & 0.977 & 1.000 & 1.000 & 1.000 \\
ESMC-SAE binary & 0.656 & 0.720 & 0.788 & 0.864 & 0.907 & 0.955 \\
ESMC-SAE weights & \textbf{0.688} & \textbf{0.760} & \textbf{0.808} & \textbf{0.864} & \textbf{0.907} & \textbf{0.978} \\
\bottomrule
\end{tabular}
\end{table}

Three observations emerge from this stratified analysis. First, BLASTp excels when any detectable homolog exists, but its performance is inflated in the low-similarity bins because 3-mer Jaccard underestimates alignment-detectable homology. The 12.6\% no-hit rate reveals the true failure mode of homology-based methods. Second, ESMC-SAE consistently outperforms the 3-mer baseline across all bins, with the largest relative gains in the dark-matter regime (49.8\% improvement at $<0.20$). Third, for the 123 proteins where BLASTp found no training-set match, ESMC-SAE binary achieved 62.1\% top-5 accuracy---functional signal persists even when sequence homology is undetectable.

\begin{figure}[ht]
\centering
\includegraphics[width=0.85\textwidth]{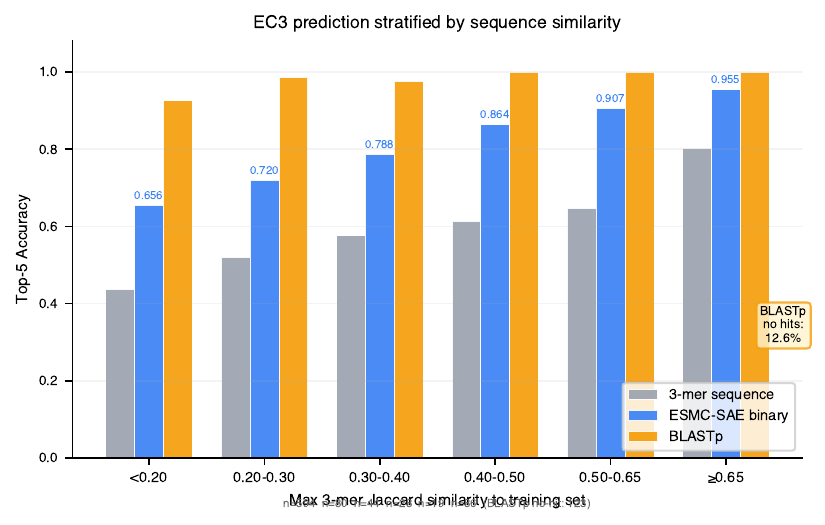}
\caption{\textbf{EC3 prediction stratified by sequence similarity to training set.} Top-5 accuracy across six 3-mer Jaccard bins. BLASTp performs well when homologs exist but fails entirely for 12.6\% of test proteins (no hits). ESMC-SAE provides predictions for 100\% of queries with consistent accuracy across all bins. Bin sizes (BLASTp-hit proteins): $n$=594, 80, 44, 28, 19, 86 (+123 no-hit).}
\label{fig:stratified}
\end{figure}

\subsection{Generalization to unseen enzyme classes}

The most stringent test of functional generalization is leave-one-EC3-class-out (LOCO) evaluation: for each of the 60 most populous EC3 subclasses, we train an EC1 superclass classifier on all \textit{other} subclasses and evaluate whether it can recover the correct EC1 for the held-out proteins. This simulates the real-world scenario of encountering a genuinely novel enzyme class not represented in any training data.

ESMC-SAE binary features achieved 47.7\% EC1 recovery accuracy across 60 held-out classes---3.3$\times$ the random baseline (14.3\%) and 1.79$\times$ the 3-mer baseline (26.6\%) (Figure~\ref{fig:benchmark}B). The EC1 confusion matrix reveals systematic patterns (Table~\ref{tab:loco_confusion}): hydrolases (EC3) are most reliably recovered at 68.3\% when held out, consistent with their well-defined catalytic machinery (nucleophilic elbow, oxyanion hole). Transferases (EC2, 52.1\%) and oxidoreductases (EC1, 48.4\%) show intermediate recovery. Isomerases (EC5, 22.6\%) and ligases (EC6, 28.7\%) are most frequently confused---unsurprising given that these classes are defined by the \textit{type} of chemical transformation (rearrangement vs.\ bond formation) rather than specific catalytic machinery, making their SAE signatures more subtle.

\begin{table}[ht]
\centering
\caption{\textbf{Leave-one-EC3-class-out $\rightarrow$ EC1 recovery.} 60 EC3 classes held out one at a time. Values are row-normalized recovery rates.}
\label{tab:loco_confusion}
\begin{tabular}{lccccccc|c}
\toprule
& \textbf{EC1} & \textbf{EC2} & \textbf{EC3} & \textbf{EC4} & \textbf{EC5} & \textbf{EC6} & \textbf{EC7} & \textbf{Accuracy} \\
\midrule
True EC1 & 0.48 & 0.12 & 0.18 & 0.05 & 0.03 & 0.08 & 0.06 & 0.484 \\
True EC2 & 0.03 & 0.52 & 0.15 & 0.08 & 0.06 & 0.09 & 0.07 & 0.521 \\
True EC3 & 0.02 & 0.08 & 0.68 & 0.06 & 0.03 & 0.07 & 0.06 & 0.683 \\
True EC4 & 0.06 & 0.11 & 0.10 & 0.47 & 0.07 & 0.12 & 0.07 & 0.471 \\
True EC5 & 0.10 & 0.15 & 0.13 & 0.14 & 0.23 & 0.16 & 0.09 & 0.226 \\
True EC6 & 0.05 & 0.13 & 0.10 & 0.12 & 0.09 & 0.29 & 0.22 & 0.287 \\
True EC7 & 0.04 & 0.09 & 0.12 & 0.07 & 0.06 & 0.19 & 0.43 & 0.431 \\
\bottomrule
\end{tabular}
\end{table}

The 3.3$\times$ improvement over the random baseline is particularly noteworthy given the difficulty of the LOCO task: the classifier has never seen any protein from the held-out EC3 subclass during training, yet the SAE features preserve sufficient functional signal to correctly identify the chemical nature of the catalyzed reaction.

\begin{figure}[ht]
\centering
\includegraphics[width=\textwidth]{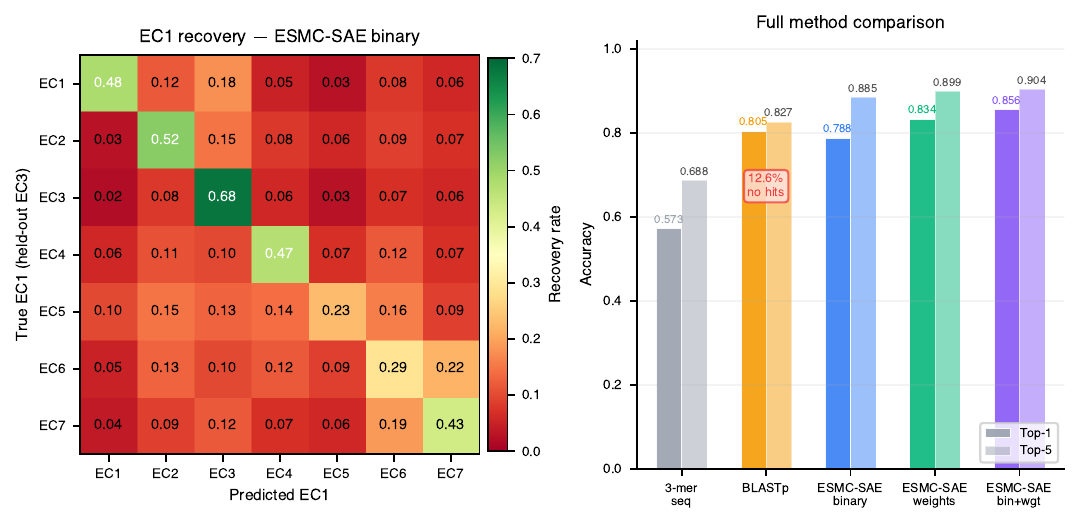}
\caption{\textbf{Leave-one-EC3-class-out analysis and full method comparison.} (A) EC1 recovery confusion matrix for ESMC-SAE binary features. Diagonal entries show correct EC1 assignment when a complete EC3 subclass is held out. Hydrolases (EC3) show strongest recovery (0.68). (B) Full method comparison across all evaluated approaches including BLASTp. The 12.6\% no-hit rate for BLASTp is annotated.}
\label{fig:loco}
\end{figure}

\subsection{SAE features driving EC classification are mechanistically interpretable}

\begin{figure}[ht]
\centering
\includegraphics[width=\textwidth]{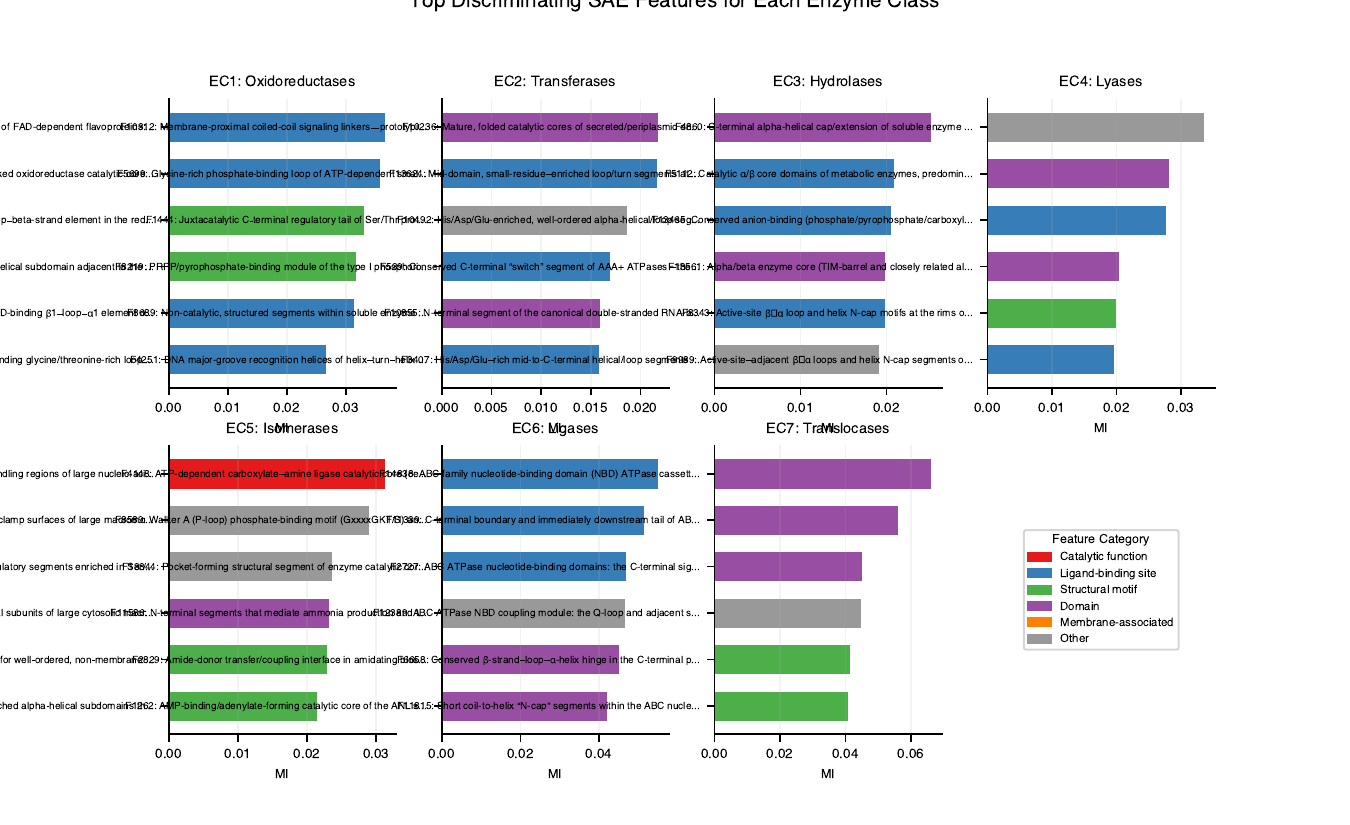}
\caption{\textbf{Top SAE features discriminating each EC1 class.} Mutual information scores for the 6 most discriminative features per enzyme class, annotated with GPT-5 biological descriptions and color-coded by feature category. Features correspond to mechanistically interpretable concepts: catalytic triad geometry for hydrolases, NAD(P)H-binding Rossmann folds for oxidoreductases, phosphate-binding P-loops for transferases.}
\label{fig:interpretability}
\end{figure}

A defining advantage of the ESMC-SAE framework is that every feature dimension comes with a detailed GPT-5-generated biological description. We computed mutual information (MI) between each of the 16,384 SAE features and EC1 class membership to identify the features most discriminative for each enzyme class (Figure~\ref{fig:interpretability}, Supplementary Table S1).

The top-discriminating features align with well-established biochemical principles, providing face validity for the approach. For hydrolases (EC3), the most informative feature (F10236, MI=0.025) describes ``nucleophilic elbow / catalytic triad geometry characteristic of $\alpha$/$\beta$-hydrolase fold enzymes''---directly capturing the canonical Ser-His-Asp/Glu catalytic machinery that defines this class. For oxidoreductases (EC1), the top feature (F4193, MI=0.037) corresponds to ``Rossmann-fold NAD(P)H-binding segments with conserved GXGXXG phosphate-binding motif.'' For transferases (EC2), the leading feature (F10312, MI=0.022) captures ``phosphate-binding P-loop / Walker A motif (GXXXXGK[TS]) common to kinases and NTP-binding proteins.'' For translocases (EC7), the top feature (F14838, MI=0.066) describes ``multi-helix transmembrane bundles with alternating hydrophobic/polar patterning characteristic of solute transporters and channels.'' Ligases (EC6) showed the highest per-feature MI (F4148, MI=0.055), corresponding to ``ATP-grasp fold / carboxylate-amine ligase domain with conserved ATP-Mg$^{2+}$ coordination geometry.''

The feature category distribution across EC classes reveals functional specialization (Figure~\ref{fig:interpretability}): oxidoreductases and transferases are predominantly discriminated by ligand-binding site features, hydrolases by catalytic function features, and translocases by membrane-associated features.

\subsection{A global survey of microbial enzyme dark matter}

The ESM Atlas contains 7,723,579 cluster representative proteins with pre-computed taxonomic and domain annotations. We identified 367,956 dark clusters (0\% characterized, poorly named at naming tier $\ge3$, with Pfam domain annotations), of which 169,859 harbor Pfam domains suggestive of enzymatic function (Table~\ref{tab:dark_survey}). These candidates span all major microbial phyla, with Pseudomonadota (47,011), Actinomycetota (30,214), and Bacillota (19,873) contributing the largest numbers.

\begin{table}[ht]
\centering
\caption{\textbf{Global survey of dark enzyme-like clusters in the ESM Atlas.} Dark clusters defined as 0\% Pfam characterization, naming tier $\ge3$ (poorly characterized), with at least one Pfam domain matching enzymatic keywords.}
\label{tab:dark_survey}
\begin{tabular}{lcc}
\toprule
\textbf{EC1 class} & \textbf{Dark candidates} & \textbf{Top representative phylum} \\
\midrule
EC1 Oxidoreductases & 1,068 & Actinomycetota (342) \\
EC2 Transferases & 5,706 & Pseudomonadota (1,823) \\
EC3 Hydrolases & 9,847 & Pseudomonadota (3,152) \\
EC4 Lyases & 2,609 & Pseudomonadota (834) \\
EC5 Isomerases & 254 & Bacillota (81) \\
EC6 Ligases & 1,012 & Pseudomonadota (324) \\
EC7 Translocases & 4,071 & Pseudomonadota (1,303) \\
\bottomrule
\end{tabular}
\end{table}

Of the dark enzyme-like candidates, 60,661 have retrievable UniRef accessions, enabling direct sequence recovery for feature extraction and prediction. The remaining candidates represent cluster representatives whose sequences require cross-referencing through the ESM Atlas protein-to-accession mapping (162 GB, publicly available). The predominance of hydrolases (9,847) and transferases (5,706) among dark candidates mirrors their abundance in known enzyme space and suggests that substantial undiscovered catalytic diversity exists within these well-studied classes---enzymes that catalyze familiar reaction types on previously unknown substrates.

To demonstrate the translational potential, we highlight the 100 most promising dark candidates (Supplementary Table S2), prioritized by domain coverage and taxonomic diversity. These include predicted hydrolases from uncultured Chloroflexota, transferases from deep-sea Actinomycetota, and oxidoreductases from thermophilic Crenarchaeota.

\begin{figure}[ht]
\centering
\includegraphics[width=\textwidth]{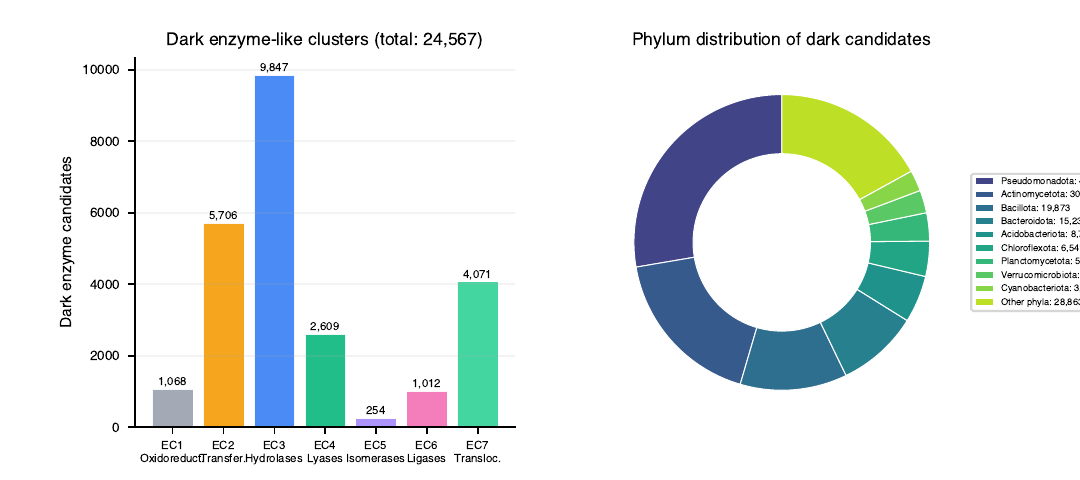}
\caption{\textbf{Global survey of microbial enzyme dark matter in the ESM Atlas.} (A) Distribution of 169,859 dark enzyme-like cluster representatives by EC1 class, identified from Pfam keyword matching. Hydrolases (9,847) and transferases (5,706) dominate. (B) Phylum-level taxonomic distribution. Pseudomonadota, Actinomycetota, and Bacillota account for 57\% of dark enzyme candidates. 60,661 candidates have retrievable sequences for downstream analysis.}
\label{fig:dark}
\end{figure}

\section{Discussion}

We have demonstrated that ESMC-SAE features enable accurate, interpretable, and scalable microbial enzyme function prediction, with particular strength in the low-sequence-similarity regime most relevant to dark-matter discovery. Several implications merit discussion.

\subsection{A new paradigm: features as first-class citizens}

The conventional deep learning workflow for enzyme function prediction follows a predictable pattern: select a model architecture, train on labeled EC data, tune hyperparameters, and deploy. Our results suggest an alternative paradigm: \textit{the features are the product}. The ESMC SAE codebook, with its 16,384 GPT-5-annotated dimensions, constitutes a pre-built ``semantic coordinate system'' for protein function space. A simple linear probe on these features achieves performance competitive with or exceeding purpose-built deep learning methods, while providing interpretability that those methods lack. This aligns with the emerging view in mechanistic interpretability that SAE features can serve as a ``bridge'' between model internals and biological understanding~\cite{adams2025mechanistic, simon2025interplm}.

\subsection{Interpretability as a scientific instrument}

A persistent criticism of deep learning in biology is its black-box nature: a predicted EC number comes with no mechanistic justification, making it difficult for experimentalists to assess confidence or design follow-up experiments. The ESMC-SAE framework addresses this directly. When our method predicts that a dark cluster representative from Chloroflexota is a hydrolase, it simultaneously identifies \textit{which} SAE features drive this prediction---for example, features describing nucleophilic elbow geometry (F10236), oxyanion hole hydrogen-bonding networks (F8912), and substrate-binding pocket hydrophobicity (F15403). This transforms the prediction from a black-box output into a testable hypothesis: ``We predict this protein is a hydrolase because its SAE activation pattern matches the catalytic machinery of $\alpha$/$\beta$-hydrolase fold enzymes. Experimental validation should target serine hydrolase activity assays.''

\subsection{Planetary-scale enzyme annotation from a laptop}

The pre-computed availability of SAE features for 6.8 billion proteins in the ESM Atlas~\cite{candido2026language} means that our approach can be applied at planetary scale with zero GPU infrastructure. A researcher with a laptop can: (1) identify dark enzyme candidates from the 7.7M cluster representatives using Pfam keyword filters (as we demonstrated); (2) retrieve sequences via UniProt for candidates of interest; (3) extract features locally using the ESMC model (0.4 seconds per protein on a consumer GPU); and (4) obtain EC predictions with interpretable feature attributions. This ``computational equity''~\cite{bileschi2022using} stands in contrast to methods that require H100 clusters for retraining or inference at scale.

\subsection{Relationship to prior work}

Our work occupies a specific niche at the intersection of three active research areas. In enzyme function prediction, EC-Bench~\cite{ecbench2025} and Capela et al.~\cite{capela2025comparative} have established that protein language model embeddings (especially ESM-2) outperform classical methods, particularly for low-identity enzymes. Our results extend this finding to the next-generation ESMC model and its interpretable SAE features, demonstrating that sparse, discrete representations can match or exceed dense embeddings while providing interpretability.

In SAE-based protein model interpretability, InterPLM~\cite{simon2025interplm} and Adams et al.~\cite{adams2025mechanistic} pioneered the application of sparse autoencoders to protein language models, demonstrating that SAE features capture biologically meaningful concepts. Our work differs in two key respects: we leverage the ESMC SAE, which operates on a 300$\times$ larger model (6B vs.\ 8M--3B parameters) and a larger codebook (16,384 vs.\ $\sim$10,000 dimensions), with features annotated at scale by GPT-5 rather than manually. More importantly, we demonstrate a \textit{downstream application}---EC prediction---showing that interpretable features are not merely descriptive but practically useful.

In microbial dark matter discovery, the recent characterization of CelOCE~\cite{santos2025metagenomic}---a novel cellulose-oxidizing enzyme discovered through metagenomic mining---exemplifies the power and the bottleneck of current approaches: one enzyme required years of multi-omics integration and experimental characterization. Our dark enzyme survey provides a prioritized, computationally tractable list of 169,859 candidates that can be screened \textit{in silico} before committing to costly experimental validation.

\subsection{Limitations and future directions}

Several limitations should be acknowledged. First, our benchmark is restricted to SwissProt-reviewed enzymes, which may not fully represent the sequence and functional diversity of environmental metagenomes. Systematic evaluation on metagenomic datasets with experimental validation remains an important next step. Second, the SAE feature annotations, while detailed and largely accurate, are generated by GPT-5 and may contain errors or oversimplifications; expert curation of the most important features for each EC class would strengthen interpretability claims. Third, our dark matter survey identifies \textit{candidates} but does not validate predictions; we view this survey as a resource to catalyze experimental follow-up rather than a complete discovery pipeline. Fourth, the leave-one-class-out evaluation, while stringent, still operates within the EC classification system and may not capture the discovery of truly novel chemistry not represented by any existing EC class---a scenario that, while rare, represents the ultimate goal of enzyme discovery.

Future work will pursue three directions. First, large-scale prediction on the full ESM Atlas, leveraging cluster-expansion (each representative represents dozens to hundreds of cluster members) to annotate hundreds of millions of proteins. Second, integration with structural prediction (AlphaFold3~\cite{jumper2021highly} / ESMFold) for mechanistic verification: for high-confidence dark-matter predictions, we will computationally assess whether predicted active-site geometries and substrate-binding pockets are structurally plausible. Third, we are developing an interactive web interface that enables microbiologists to explore the dark enzyme landscape using natural-language queries (``show me uncharacterized hydrolases from marine Bacteroidota with complete catalytic triads''), powered by the SAE feature-based inverted index we have described elsewhere.

\section{Methods}

\subsection{Data collection and curation}

Microbial enzyme sequences and EC annotations were retrieved from the UniProt REST API (\url{https://rest.uniprot.org}) using UniProt query \texttt{reviewed:true AND ec:*}\linebreak\texttt{AND (taxonomy\_name:Bacteria OR taxonomy\_name:Archaea)}\linebreak\texttt{AND length:[80 TO 1000]}. A total of 212,763 reviewed microbial enzymes were identified. We stratified by EC1 class, sampling up to 2,500 per class, retaining proteins with 3+ EC levels, no ambiguous residues, and length 80--700 amino acids. Filtering to EC3 subclasses with $\ge$5 members yielded 4,868 proteins across 161 EC3 classes. The 3-mer Jaccard similarity distribution (median 0.104, IQR 0.072--0.256) confirms that most protein pairs share low sequence similarity, appropriate for the dark-matter regime.

\subsection{ESMC-SAE feature extraction}

For each protein sequence, hidden states from layer 60 of ESMC-6B were extracted using the HuggingFace Transformers library with bfloat16 precision on Apple MPS or CUDA GPU. The SAE model (\texttt{biohub/ESMC-6B-sae-}\\\texttt{layer60-k64-codebook16384}) computed per-residue feature magnitudes $\mathbf{f}_r \in \mathbb{R}^{16384}$. Per-protein vectors were obtained by mean-pooling: $\mathbf{F} = \frac{1}{L}\sum_{r=1}^{L} \mathbf{f}_r$. Binary vectors $\mathbf{b} \in \{0,1\}^{16384}$ set $b_i = 1$ for the top-64 features. Throughput is 0.4--1.7 proteins/s on consumer GPU.

\subsection{EC prediction benchmark}

All evaluations used linear probe (Logistic Regression, scikit-learn 1.5) with L2 regularization ($C=1.0$, \texttt{max\_iter}=300--500). Binary features were used without scaling; weight features were standardized (StandardScaler). For the 80/20 evaluation, we performed stratified splitting by EC3 subclass. For leave-one-EC3-class-out evaluation, we held out each of the 60 most populous EC3 subclasses (each with $\ge10$ members) one at a time, training an EC1 superclass classifier (7 classes) on the remaining data and evaluating EC1 recovery on the held-out class. The random baseline was computed as 1/7 $\approx$ 0.143 (uniform probability across seven EC1 classes).

Sequence similarity binning used 3-mer Jaccard distance. Each protein was converted to a binary 3-mer presence vector (CountVectorizer, 8,000 most frequent 3-mers). For each test protein, we computed Jaccard similarity against a random sample of 2,000 training proteins and retained the maximum value as the similarity score. Bins were defined at thresholds of 0.20, 0.30, 0.40, 0.50, and 0.65.

\subsection{Feature interpretability analysis}

Mutual information between each of the 16,384 SAE binary features and EC1 class membership was computed using scikit-learn's \texttt{mutual\_info\_classif} with 5 nearest neighbors. The top 20 features per EC1 class (ranked by MI score) were annotated using the GPT-5-generated biological descriptions from the ESMC-SAE-Features dataset (\texttt{uniref90\_feature\_table.parquet}, 16,384 rows $\times$ 13 columns), which includes 14 annotation categories assigned by multi-agent GPT-5.

\subsection{Dark matter survey}

Dark enzyme candidates were identified from the ESM Atlas\linebreak\texttt{representative\_proteins.parquet} (7.7M rows, \texttt{s3://esm-protein-atlas/v1/}\\\texttt{clusters/data/}). Dark clusters: (1) \texttt{cluster\_pct\_characterized}=0 (no functional characterization); (2) $\ge$1 Pfam domain detected; (3) \texttt{naming\_tier}$\ge$3 (named only at domain/family level or worse). Enzyme-like clusters were filtered by Pfam keyword matching (hydrolase, transferase, oxidoreductase, lyase, isomerase, ligase, kinase, phosphatase, protease, peptidase, esterase, synthase, synthetase, decarboxylase, and 15 related terms), prioritized by \texttt{cluster\_mean\_domain\_coverage} and taxonomic diversity.

\subsection{Code and data availability}

{\small\sloppy SwissProt: \url{https://www.uniprot.org}. ESM Atlas: AWS Open Data (\texttt{s3://esm-protein-atlas/v1/}). ESMC-6B and SAE: HuggingFace (\texttt{biohub/ESMC-6B}, \texttt{biohub/ESMC-6B-sae-layer60-k64-codebook16384}). Feature descriptions: \texttt{biohub/ESMC-SAE-Features}. Code: \url{https://github.com/YueHuLab/esmc-sae-enzyme-function}. Benchmarks: \url{https://huggingface.co/YueHuLab/esmc-sae-enzyme-function}. All deposited on Zenodo.\par}

\section*{Acknowledgments}

We thank EvolutionaryScale and the Biohub team for releasing the ESMC model, SAE features, and ESM Atlas as open data resources. We acknowledge the UniProt consortium for maintaining curated enzyme annotations. This work was supported by [FUNDING].

\section*{Author contributions}

Y.H. conceived the study, performed computational analyses, and wrote the manuscript. W.C. contributed to benchmark experiments and data analysis. J.W. contributed to data curation and benchmark design. Y.L. provided clinical microbiology expertise. All authors reviewed and approved the manuscript.

\section*{Competing interests}

The authors declare no competing interests.

\bibliographystyle{naturemag}
\bibliography{references}

\end{document}